\documentclass{PoS}

\usepackage[utf8]{inputenc}
\usepackage{amsmath}
\usepackage{amsfonts}
\usepackage{amssymb}
\usepackage{bbold}
\usepackage{tikz}
\usepackage{braket}
\usetikzlibrary{positioning,shapes,shadows,arrows,decorations,decorations.pathmorphing,calc}

\definecolor{Green}{RGB}{0,128,64}
\definecolor{Cyan}{RGB}{0,128,128}

\newcommand{\Op}{\mathcal{O}}

\newcommand{\ME}{\mathcal{M}}

\newcommand{\dd}{\mathrm{d}}

\newcommand{\pvec}{\mathbf{p}}
\newcommand{\xvec}{\mathbf{x}}
\newcommand{\yvec}{\mathbf{y}}

\newcommand{\rvec}{\mathbf{r}}

\newcommand{\nvec}{\mathbf{0}}
\newcommand{\ie}{i.e.\ }
\newcommand{\eg}{e.g.\ }

\newcommand{\wrt}{w.r.t.\ }
\newcommand{\ts}[1]{\textstyle #1 \displaystyle}

\newcommand{\fref}[1]{Figure \ref{#1}}
\newcommand{\tref}[1]{Table \ref{#1}}
\newcommand{\sref}[1]{Section \ref{#1}}

\title{Double Parton Distributions of the Pion}

\ShortTitle{DPDs of the Pion}

\author{\speaker{Christian Zimmermann} \thanks{for RQCD}\\
        University of Regensburg\\
        E-mail: \email{christian.zimmermann@physik.uni-regensburg.de}}

\abstract{The effects of double hard interactions are no longer negligible at energy scales reached at the LHC. Double parton scattering (DPS) processes are often described by taking the product of two single parton scattering processes assuming that interference effects are very small. We calculate four point functions (4pt-functions), which appear in the the DPS cross section, employing lattice techniques. We consider a pion at rest and test the validity of the afore-mentioned factorization assumption by convoluting two pion form factors and comparing the result to the 4pt data. For our calculations we use a $N_f = 2$ gauge ensemble on a $40^3 \times 64$ lattice, with lattice spacing $a = 0.071 \mathrm{fm}$ and pion mass $m_{\pi} = 288.8 \mathrm{MeV}$.}

\FullConference{34th annual International Symposium on Lattice Field Theory\\
		24-30 July 2016\\
		University of Southampton, UK}

\begin{document}

\section{Introduction}
\label{sec:Introduction}
Hadron-hadron collisions are an important subject when discovering new physics at high energy scales. Therefore the correct description of the underlying event in these interactions is essential. At energy scales reached in LHC, multi parton interactions (MPIs) are no longer negligible. The most important MPI processes are double parton scattering (DPS) processes, which are often naively described by neglecting any interference effects:
\begin{align}
\dd \sigma_{\mathrm{DPS}} = \frac{\dd \sigma_{\mathrm{SPS}} \dd \sigma_{\mathrm{SPS}}}{C \sigma_{\mathrm{eff}}}\ ,
\label{pocket}
\end{align}
where $\sigma_{\mathrm{SPS}}$ is the single parton scattering (SPS) cross section, and $\sigma_{\mathrm{eff}}$ is some effective cross section, which is related with the hadron size. $C=2$ or $1$ denotes some symmetry factor \cite{Calucci:1999yz}.\\
A more fundamental description can be done by introducing so-called double parton distributions (DPDs), which involve two-operator matrix elements.
Our intention is to obtain these matrix elements from calculations of 4pt-functions on the lattice, starting with the pion at rest. We also want to check to what extent the "naive factorization" ansatz \eqref{pocket}  is valid by comparing the convolution of two form factors with our lattice results for the 4pt-functions.

\section{Double Parton Scattering and Double Parton Distributions}
\label{sec:DPS_DPD}
For the description of a double parton scattering (DPS) process, it is usually assumed that the process factorizes into a soft and a hard part. In that case, one can write the cross section as (in the following lightcone coordinates are used):
\begin{align}
\begin{aligned}
\frac{\dd \sigma}{\dd x_1 \dd \bar{x}_1 \dd x_2 \dd \bar{x}_2} &= \sum_{\substack{\mathrm{ polarization}\\ \mathrm{flavor}}} \frac{\sigma_1 \sigma_2}{C} \int \dd^2 \yvec_{\perp} F(x_i, \yvec_{\perp}) F(\bar{x}_i,\yvec_{\perp})\ .
\end{aligned}
\end{align}
A derivation from first principles is given in \cite{Die11}. In the equation above the $\sigma_i$ denote the parton scattering cross sections. The variable $y$ can be interpreted as the distance between the scattering partons. $F(x_i,\yvec_{\perp})$ are the so-called collinear double parton distributions (DPDs), which involve hadronic matrix elements of two lightcone operators. These contain Wilson lines, once we include higher order contributions.\\
This is different for Mellin moments of DPDs, where the two parton momentum fractions are integrated out, which has the consequence that the operators become local and therefore Wilson lines do not appear anymore. Lattice calculations become more feasible in this case. Explicitly one finds for the lowest moment:
\begin{align}
M^p(y) &= \frac{(p^+)^{-1}}{2} \int \dd y^- \ME^p(y) \label{dpd_mellin_mom} \\
\ME^p(y) &= \left. \bra{h(p)} \Op^{f_1 f^\prime_1}_1(0) \Op^{f_2 f^\prime_2}_2(y) \ket{h(p)} \right|_{y^+ = 0} \label{local_me} \\ 
\Op^{f f^\prime}_{1,2}(y) &= \bar{q}^{f}(y) \Gamma q^{f^\prime}(y)\ , \label{local_ops}
\end{align}
where $h(p)$ denotes a specific hadron with momentum $p$. $\Gamma$ is a combination of Dirac matrices, which depends on the polarization of the quarks taking part on the interaction. We will only discuss the vector and axial vector case and furthermore the scalar and pseudoscalar channels, although the latter two correspond to higher twist contributions.\\
In the following we will use decompositions of the matrix elements \wrt their Lorentz structure into invariant functions, \eg :
\begin{align}
\ME^p_{\mathrm{SS}/\mathrm{PP}}(y) = 2m^2 A_{\mathrm{SS}/\mathrm{PP}}(py,y^2) \qquad \qquad \qquad \qquad \qquad \qquad \qquad \qquad \qquad \qquad \ \ \, \, \label{ASS} \\
\begin{aligned}
\mathcal{T} \ME^{p,\{\mu\nu \}}_{\mathrm{VV}}(y) &= \left[ 2p^\mu p^\nu - \ts{\frac{1}{2}} g^{\mu\nu} p^2 \right] A_{\mathrm{VV}}(py,y^2) + m^2 \left[ 2p^{\{ \mu} y^{\nu\} } - \ts{\frac{1}{2}} g^{\mu\nu} py \right] B_{\mathrm{VV}}(py,y^2)\\
&+ m^4 \left[ 2y^\mu y^\nu - \ts{\frac{1}{2}} g^{\mu\nu} y^2 \right] C_{\mathrm{VV}}(py,y^2) \ ,
\end{aligned}
\label{AVV}
\end{align}
where $\mathcal{T}$ denotes trace subtraction and the $\{ \}$-notation indicates symmetrization \wrt Lorentz indices.\\
As mentioned before we intend to check the validity of the factorization assumption \eqref{pocket}, \ie the factorization of a two-operator matrix element into two one-operator matrix elements. It seems natural that a suitable way is to insert a complete set of intermediate states, assuming that states of the observed hadron dominate:
\begin{align}
\bra{h(p)} \Op_1 \Op_2 \ket{h(p)} \approx \int \frac{\dd^4 p^{\prime}}{(2\pi)^4} \bra{h(p)} \Op_1 \ket{h(p^\prime)} \bra{h(p^\prime)} \Op_2 \ket{h(p)} \delta(p^{\prime 2} - m^2)\ .
\label{fact_ansatz}
\end{align}
This ansatz one can apply to lightcone matrix elements as well as to matrix elements of local operators. We will check both possibilities in \sref{sec:fact}.\\

\section{DPDs from Lattice Studies}
\label{sec:DPD_latt}
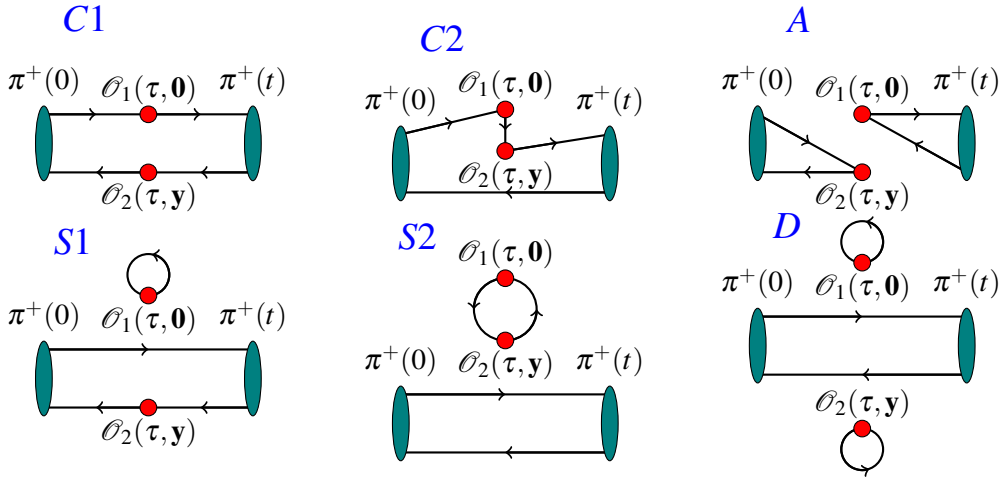
\begin{figure}
\begin{center}
\begin{minipage}[c][3cm]{4.6cm}
\begin{tikzpicture}[scale=0.55]
\draw[thick] (0,0.7) -- (5,0.7);
\draw[thick] (0,-0.7) -- (5,-0.7);
\draw[thick,->] (0,0.7) -- (1.25,0.7);
\draw[thick,->] (2.5,0.7) -- (3.75,0.7);
\draw[thick,->] (2.5,-0.7) -- (1.25,-0.7);
\draw[thick,->] (5,-0.7) -- (3.75,-0.7);
\filldraw[fill=Cyan] (0,0) circle [x radius=0.2, y radius=0.9];
\draw (0,1.5) node {$\pi^+(0)$};
\filldraw[fill=Cyan] (5,0) circle [x radius=0.2, y radius=0.9];
\draw (5,1.5) node {$\pi^+(t)$};
\filldraw[fill=red] (2.5,0.7) circle [radius=0.2] node [above] {$\Op_1(\tau,\nvec)$};
\filldraw[fill=red] (2.5,-0.7) circle [radius=0.2] node [below] {$\Op_2(\tau,\yvec)$};
\draw (1,3) node {\Large \color{blue} $C1$};
\end{tikzpicture}
\end{minipage}
\begin{minipage}[c][3cm]{4.6cm}
\begin{tikzpicture}[scale=0.55]
\draw[thick] (0,0.7) -- (2.5,1.3) -- (2.5,0.3) -- (5,0.7);
\draw[thick] (0,-0.7) -- (5,-0.7);
\draw[thick,->] (0,0.7) -- (1.25,1);
\draw[thick,->] (2.5,1.3) -- (2.5,0.8);
\draw[thick,->] (2.5,0.3) -- (3.75,0.5);
\draw[thick,->] (5,-0.7) -- (2.5,-0.7);
\filldraw[fill=Cyan] (0,0) circle [x radius=0.2, y radius=0.9];
\draw (0,1.5) node {$\pi^+(0)$};
\filldraw[fill=Cyan] (5,0) circle [x radius=0.2, y radius=0.9];
\draw (5,1.5) node {$\pi^+(t)$};
\filldraw[fill=red] (2.5,1.3) circle [radius=0.2] node [above] {$\Op_1(\tau,\nvec)$};
\filldraw[fill=red] (2.5,0.3) circle [radius=0.2] node [below] {$\Op_2(\tau,\yvec)$};
\draw (1,3) node {\Large \color{blue} $C2$};
\end{tikzpicture}
\end{minipage}
\begin{minipage}[c][3cm]{4.6cm}
\begin{tikzpicture}[scale=0.55]
\draw[thick] (0,0.7) -- (2.5,-0.7) -- (0,-0.7);
\draw[thick] (5,-0.7) -- (2.5,0.7) -- (5,0.7);
\draw[thick,->] (0,0.7) -- (1.25,0);
\draw[thick,->] (2.5,-0.7) -- (1.1,-0.7);
\draw[thick,->] (5,-0.7) -- (3.75,0);
\draw[thick,->] (2.5,0.7) -- (3.9,0.7);
\filldraw[fill=Cyan] (0,0) circle [x radius=0.2, y radius=0.9];
\draw (0,1.5) node {$\pi^+(0)$};
\filldraw[fill=Cyan] (5,0) circle [x radius=0.2, y radius=0.9];
\draw (5,1.5) node {$\pi^+(t)$};
\filldraw[fill=red] (2.5,0.7) circle [radius=0.2] node [above] {$\Op_1(\tau,\nvec)$};
\filldraw[fill=red] (2.5,-0.7) circle [radius=0.2] node [below] {$\Op_2(\tau,\yvec)$};
\draw (1,3) node {\Large \color{blue} $A$};
\end{tikzpicture}
\end{minipage}
\begin{minipage}[c][3cm]{4.6cm}
\begin{tikzpicture}[scale=0.55]
\draw[thick] (0,0.7) -- (5,0.7);
\draw[thick] (0,-0.7) -- (5,-0.7);
\draw[thick] (2.5,2) arc (-90:270:0.5);
\draw[thick,->] (2.5,2) arc (-90:80:0.5);
\draw[thick,->] (0,0.7) -- (2.5,0.7);
\draw[thick,->] (2.5,-0.7) -- (1.25,-0.7);
\draw[thick,->] (5,-0.7) -- (3.75,-0.7);
\filldraw[fill=Cyan] (0,0) circle [x radius=0.2, y radius=0.9];
\draw (0,1.5) node {$\pi^+(0)$};
\filldraw[fill=Cyan] (5,0) circle [x radius=0.2, y radius=0.9];
\draw (5,1.5) node {$\pi^+(t)$};
\filldraw[fill=red] (2.5,2) circle [radius=0.2] node [below] {$\Op_1(\tau,\nvec)$};
\filldraw[fill=red] (2.5,-0.7) circle [radius=0.2] node [below] {$\Op_2(\tau,\yvec)$};
\draw (0.7,3.3) node {\Large \color{blue} $S1$};
\end{tikzpicture}
\end{minipage}
\begin{minipage}[c][3cm]{4.6cm}
\begin{tikzpicture}[scale=0.55]
\draw[thick] (0,0.7) -- (5,0.7);
\draw[thick] (0,-0.7) -- (5,-0.7);
\draw[thick] (2.5,2) arc (-90:270:0.75);
\draw[thick,->] (2.5,2) arc (-90:0:0.75);
\draw[thick,->] (2.5,3.5) arc (90:180:0.75);
\draw[thick,->] (0,0.7) -- (2.5,0.7);
\draw[thick,->] (5,-0.7) -- (2.5,-0.7);
\filldraw[fill=Cyan] (0,0) circle [x radius=0.2, y radius=0.9];
\draw (0,1.5) node {$\pi^+(0)$};
\filldraw[fill=Cyan] (5,0) circle [x radius=0.2, y radius=0.9];
\draw (5,1.5) node {$\pi^+(t)$};
\filldraw[fill=red] (2.5,2) circle [radius=0.2] node [below] {$\Op_2(\tau,\yvec)$};
\filldraw[fill=red] (2.5,3.5) circle [radius=0.2] node [above] {$\Op_1(\tau,\nvec)$};
\draw (0.4,4.5) node {\Large \color{blue} $S2$};
\end{tikzpicture}
\end{minipage}
\begin{minipage}[c][3cm]{4.6cm}
\begin{tikzpicture}[scale=0.55]
\draw[thick] (0,0.7) -- (5,0.7);
\draw[thick] (0,-0.7) -- (5,-0.7);
\draw[thick] (2.5,2) arc (-90:270:0.5);
\draw[thick] (2.5,-2) arc (90:450:0.5);
\draw[thick,->] (2.5,2) arc (-90:80:0.5);
\draw[thick,->] (0,0.7) -- (2.5,0.7);
\draw[thick,->] (2.5,-2) arc (90:290:0.5);
\draw[thick,->] (5,-0.7) -- (2.5,-0.7);
\filldraw[fill=Cyan] (0,0) circle [x radius=0.2, y radius=0.9];
\draw (0,1.5) node {$\pi^+(0)$};
\filldraw[fill=Cyan] (5,0) circle [x radius=0.2, y radius=0.9];
\draw (5,1.5) node {$\pi^+(t)$};
\filldraw[fill=red] (2.5,2) circle [radius=0.2] node [below] {$\Op_1(\tau,\nvec)$};
\filldraw[fill=red] (2.5,-2) circle [radius=0.2] node [above] {$\Op_2(\tau,\yvec)$};
\draw (0.7,2.9) node {\Large \color{blue} $D$};
\end{tikzpicture}
\end{minipage}
\end{center}
\caption{Depiction of the six possible independent Wick contractions.}
\label{Fig:4pt-graphs}
\end{figure}

\begin{table}
\begin{center}
\begin{tabular}{cccccccc}
\hline \hline 
Ensemble & $\beta$ & $a [\mathrm{fm}]$ & $\kappa$ & $V$ & $m_\pi [\mathrm{GeV}]$ & $N(N_{4\mathrm{pt}})$ & $N_\mathrm{sm}$ \\
\hline
I & $5.29$ & $0.071$ & $0.13632$ & $40^3\times 64$ & $0.2888(11)$ & $2025(984)$ & $400$ \\
\hline \hline
\end{tabular}
\end{center}
\caption{Details of the ensemble used for the simulation. $N_{4\mathrm{pt}}$ labels the number of configurations, which are used for our 4pt-calculations. $N_{\mathrm{sm}}$ denotes the number of smearing iterations.}
\label{Tab:ens}
\end{table}

\begin{table}
\begin{center}
\begin{tabular}{cccccc}
\hline \hline
 & S & P & V & A & T \\
\hline
$Z$ & $0.4577(18)$ & $0.3538(92)$ & $0.7365(48)$ & $0.76487(64)$ & $0.9141(26)$\\
$Z_\mathrm{conv}$ & $1.3543$ & $1.3543$ & $1$ & $1$ & $0.93313$ \\
\hline \hline
\end{tabular}
\end{center}
\caption{Renormalization factors $Z$ and factors $Z_{\mathrm{conv}}$ for the conversion to the $\overline{\mathrm{MS}}$-scheme. S, P, V, A and T denote the different operator insertion types.}
\label{Tab:ren}
\end{table}

We are able to obtain some information about the matrix elements \eqref{local_me} by calculating 4pt-functions on the lattice. We start with the calculation for the pion with zero momentum taking only channels containing no derivative terms corresponding to the first Mellin moment. These matrix elements can be obtained by calculating the following ratio, in a limit, where excitations are suppressed (this is analogous to the calculation of one-operator matrix elements, see \eg \cite{Capitani:1998ff}):
\begin{align}
\ME(\yvec) = \left. 2m \frac{C_{4\mathrm{pt}}(t,\tau,\yvec)}{C_{2\mathrm{pt}}(t)} \right|_{0 \ll \tau \ll t}\ ,
\end{align}
where $C_{2\mathrm{pt}}(t)$ is the usual pion 2pt-function and $C_{4\mathrm{pt}}(t,\tau,\yvec)$ denotes the 4pt-function:
\begin{align}
C_{4\mathrm{pt}}(t,\tau,\yvec) &= \langle \Op_{\pi^+}(t) \Op_{1}(\nvec,\tau) \Op_{2}(\yvec,\tau) \Op^{\dagger}_{\pi^+}(0) \rangle \\
C_{2\mathrm{pt}}(t) &= \langle \Op_{\pi^+}(t) \Op^{\dagger}_{\pi^+}(0) \rangle\ .
\end{align}
$\Op^{\dagger}_{\pi^+}(t)$ denotes the pion interpolator, where we use:
\begin{align}
\Op^{\dagger}_{\pi^+}(t) = \frac{1}{V} \sum_{\xvec} \bar{u}(\xvec,t) \gamma_5 d(\xvec,t)\ .
\end{align}
For the calculation of the 4pt-function, we have to consider six independent Wick contractions, which are depicted in \fref{Fig:4pt-graphs}.
The two operator insertions are placed at the same time slice, which has a time separation $\tau$ to the pion source.
For the time separation between source and sink we choose $t=15a$. Hence we expect a plateau in the region $ 6 \lesssim \tau \lesssim 9$.
For the evaluation of the 4pt-graphs we use an ensemble of $N_f=2$ gauge configurations (see \tref{Tab:ens}, same as ensemble V in \cite{Bali:2014nma}). The gauge fields are smoothed using APE smearing \cite{FALCIONI1985624}. For our calculations we use the One-End-Trick incorporating stochastic $\mathbb{Z}_2 \otimes \mathbb{Z}_2$-sources \cite{Boyle:2008rh}, which are improved by Wuppertal smearing \cite{Gusken:1989qx}. 
We renormalize our results and convert it to the $\overline{\mathrm{MS}}$-scheme at a scale $\mu = 2\mathrm{GeV}$ using the renormalization and conversion factors listed in \tref{Tab:ren} (see \cite{Gockeler:2010yr}).\\
\begin{figure}
\begin{center}
\begin{minipage}[c][5cm]{6.5cm}
\includegraphics[width=6.5cm,trim={20 0 280 0},clip=true]{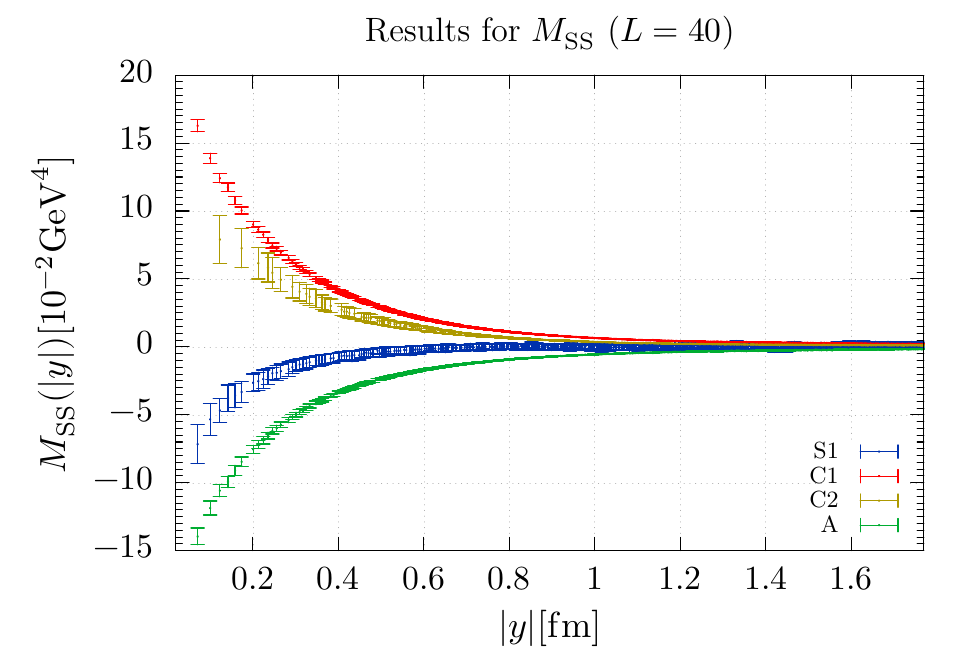}
\end{minipage}
\begin{minipage}[c][5cm]{6.5cm}
\includegraphics[width=6.5cm,trim={20 0 280 0},clip=true]{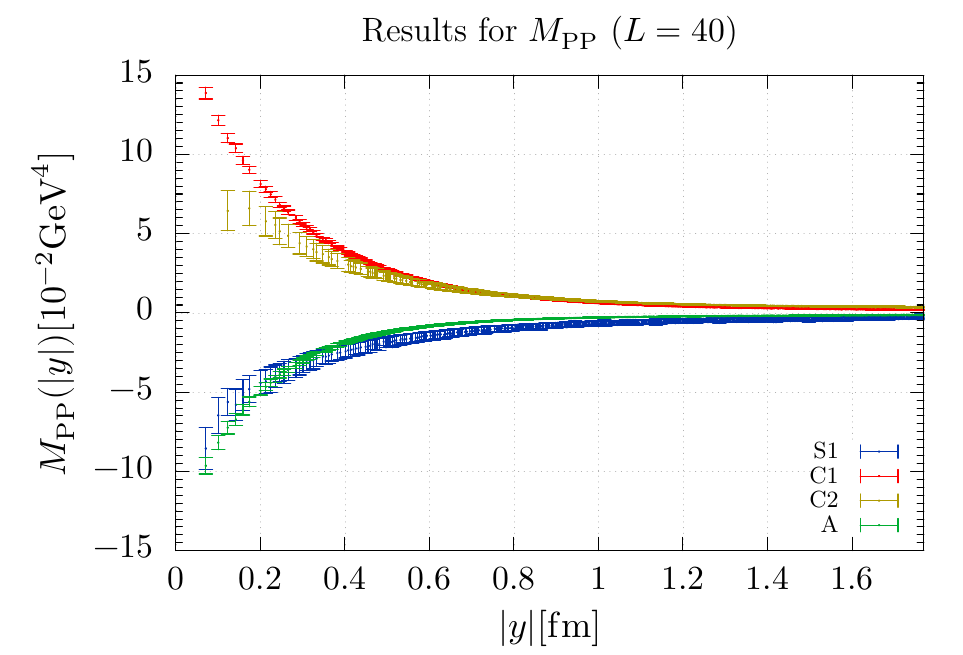}
\end{minipage}
\begin{minipage}[c][4cm]{6.5cm}
\includegraphics[width=6.5cm,trim={20 0 280 0},clip=true]{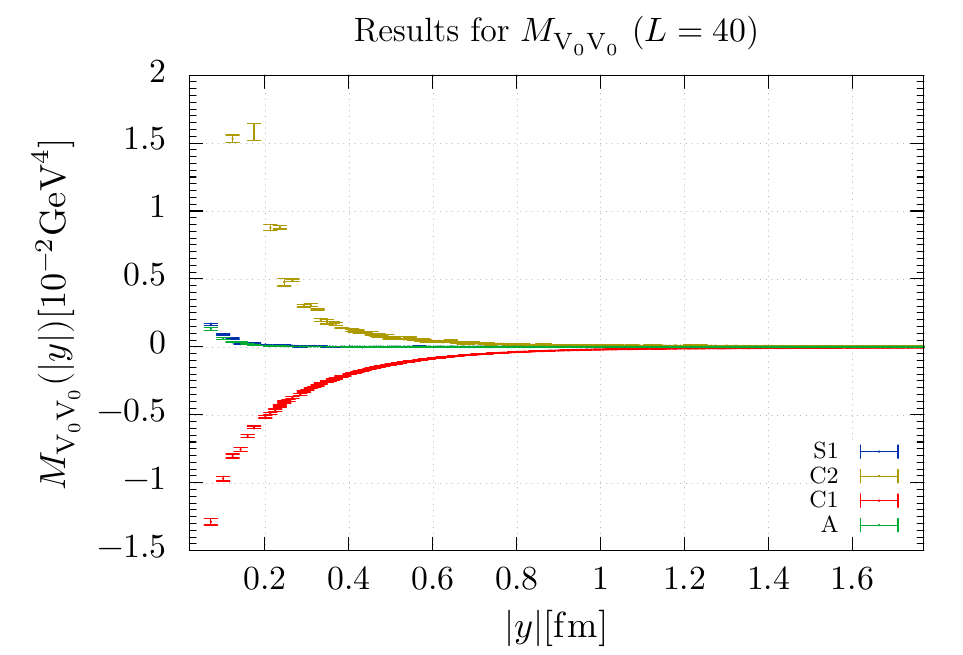}
\end{minipage}
\begin{minipage}[c][4cm]{6.5cm}
\includegraphics[width=6.5cm,trim={20 0 280 0},clip=true]{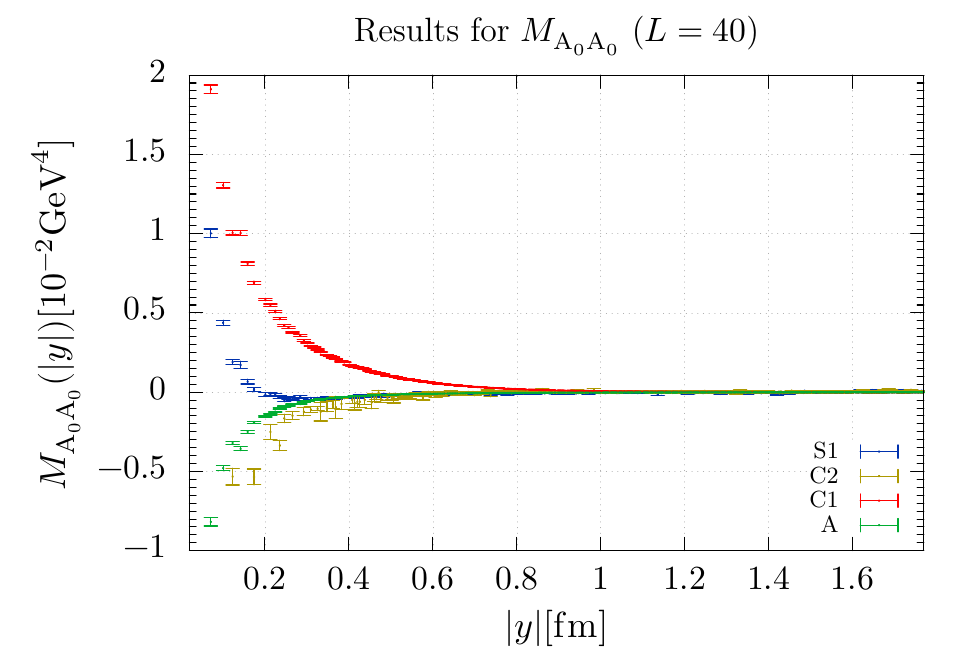}
\end{minipage}
\end{center}
\caption{Results for two-operator matrix elements for the channels SS, PP, $\mathrm{V}_0 \mathrm{V}_0$ and $\mathrm{A}_0 \mathrm{A}_0$. Here the results of the graphs C1, C2, S1 and A are shown separately.}
\label{fig:results_me}
\end{figure}

For the graphs C1, C2, S1 and A we obtain clear non-zero signals, while for the graphs S2 and D the signals are very noisy. For some matrix elements these results are shown in \fref{fig:results_me}. We are able to project out the invariant functions appearing in \eqref{ASS} and \eqref{AVV}, which are not shown here (the C1 data of the scalar and vector channels is shown in \fref{fig:test1_res} compared with the form factor convolutions). Notice that disconnected graphs are less suppressed in the (pseudo)scalar channel than in the (axial) vector channel.


\section{Naive Factorization}
\label{sec:fact}

Finally we want to check the "naive factorization" into one-operator matrix elements. Applying \eqref{fact_ansatz} to matrix elements of lightcone operators yields the following relation (test 1):
\begin{align}
A_{\mathrm{VV}/\mathrm{SS}}(py\! =\! 0, y^2) \approx \frac{\eta_C}{\pi} \int_0^1 \dd \zeta \left(1-\frac{\zeta}{2}\right)^2 \left(1-\zeta \right)^{-1} \int \frac{\dd^2 \rvec_{\perp}}{(2\pi)^2} e^{-i \yvec_{\perp}\cdot \rvec_{\perp}} F_{\mathrm{V}/\mathrm{S}}(t(\zeta, \rvec_{\perp}))\ ,
\label{ftest1}
\end{align}
where $t(\zeta, \rvec_{\perp}) = - (\zeta^2 m^2 + \rvec^2_{\perp}) / 	(1 - \zeta)$ and $\eta_C = 1$ for the scalar case and $-1$ for the vector case. On the other hand \eqref{fact_ansatz} may be inserted in a matrix element of local operators, which can be brought into the following form (test 2):
\begin{align}
\begin{aligned}
\bra{\pi^+(p)} \Op^{uu}_{\mathrm{V}_0}(\nvec) \Op^{dd}_{\mathrm{V}_0}(\yvec) \ket{\pi^+(p)} &\approx
-\int_0^\infty \frac{\dd (\rvec^2)}{4\pi^2 |\yvec|}  \frac{\sin (|\yvec||\rvec|) \left(m + E(\rvec^2)\right)^2}{2 E(\rvec^2)} F_{\mathrm{V}}^2(t(\rvec^2)) \\
\bra{\pi^+(p)} \Op^{uu}_{\mathrm{S}}(\nvec) \Op^{dd}_{\mathrm{S}}(\yvec) \ket{\pi^+(p)} &\approx
\int_0^\infty \frac{\dd (\rvec^2)}{4\pi^2 |\yvec|} \frac{\sin (|\yvec||\rvec|)}{2 E(\rvec^2)} F_{\mathrm{S}}^2(t(\rvec^2))\ ,
\end{aligned}
\label{ftest2}
\end{align}
where $E(\rvec) = \sqrt{\rvec^2 + m^2}$ and $t(\rvec) = 2m^2- 2m E(\rvec)$. Both tests involve the electromagnetic ($F_{\mathrm{V}}(t)$) or scalar form factor ($F_{\mathrm{S}}(t)$) of the pion, respectively. These we obtain from 3pt-functions on the lattice. Here we consider for the moment only connected graphs and use contributions of momenta, for which $|\pvec| \le \frac{2\pi}{L}\sqrt{3}$. Notice that both tests trivially fail for the pseudoscalar or axial vector case, since the corresponding form factors are exactly zero for symmetry reasons. This is, however, not the case for two-operator matrix elements (see \sref{sec:DPD_latt}).\\
To perform the integrals in \eqref{ftest1} and \eqref{ftest2}, we use the following parametrization of the pion form factor:
\begin{align}
F(t) = F_0 \left( 1 + \frac{t}{M^2} \right)^{-p}\ .
\end{align}
The parameters $F_0$ and $M$ may be obtained from a covariant fit on the 3pt data (in the el.\ m.\ case $F_0$ was fixed to $1$ according to charge conservation). To get an estimate of our systematic error, we perform several fits varying the fixed value of $p$ (for the el.\ m.\  form factor we also treat $p$ as a free fit parameter, since the data quality is high enough). The fit results are shown in \tref{Tab:ff_fit}.\\

\begin{table}
\begin{center}
\begin{tabular}{cccccc}
\hline \hline
 $\#$ & quantity & $F_0$ & $M [\mathrm{GeV}]$ & $p$ & $\chi^2 / \mathrm{DOF}$ \\
\hline
$1$ & $F_{\mathrm{em}}$ & $1 \mathrm{(fixed)}$ & $0.777(12)$ & $1 \mathrm{(fixed)}$ & $6.010$ \\
$2$  & & $1 \mathrm{(fixed)}$ & $0.872(16)$ & $1.173(69)$ & $4.400$ \\
\hline
$3$ & $F_{\mathrm{scal}}$ & $2.222(19)\mathrm{GeV}$ & $1.314(39)$ & $1 \mathrm{(fixed)}$ & $7.886$ \\
$4$ & & $2.212(19)\mathrm{GeV}$ & $2.023(50)$ & $2 \mathrm{(fixed)}$ & $9.877$ \\
\hline \hline
\end{tabular}
\end{center}
\caption{Results for the fits on the electromagnetic or scalar pion form factor. All fits take into account error correlations. The largest contributing value of $|t|$ in the fit is $1.143\ \mathrm{GeV}^2$.}
\label{Tab:ff_fit}
\end{table}

\begin{figure}
\begin{center}
\begin{minipage}[c][4cm]{7cm}
\includegraphics[width=7cm,trim={20 0 0 0},clip=true]{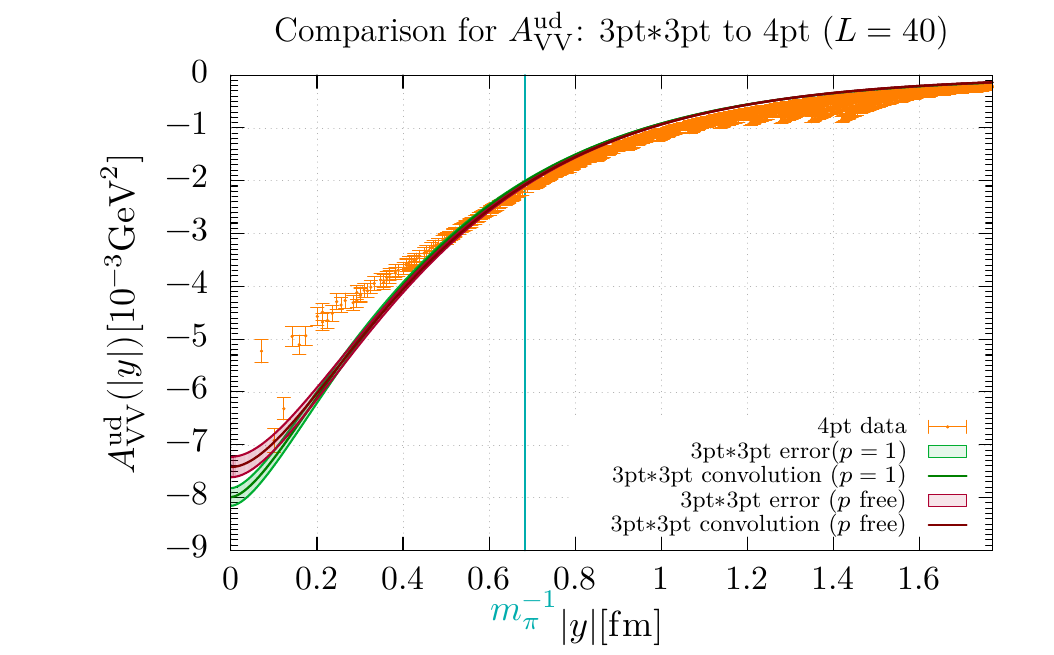}
\end{minipage}
\begin{minipage}[c][4cm]{7cm}
\includegraphics[width=7cm,trim={20 0 0 0},clip=true]{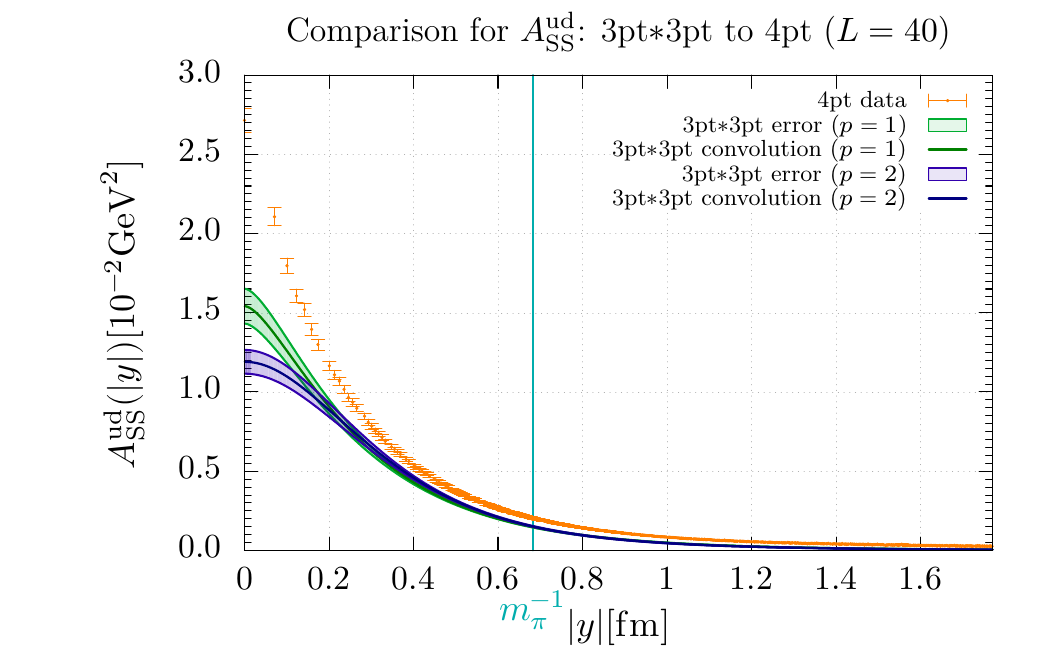}
\end{minipage}
\end{center}
\caption{Results for the first naive factorization test according to \protect\eqref{ftest1} for our two fit results. The convolutions are compared with the 4pt data for $A_{\mathrm{VV}}$ and $A_{\mathrm{SS}}$.}
\label{fig:test1_res}
\end{figure}
\begin{figure}
\begin{center}
\begin{minipage}[c][4cm]{7cm}
\includegraphics[width=7cm,trim={300 0 0 0},clip=true]{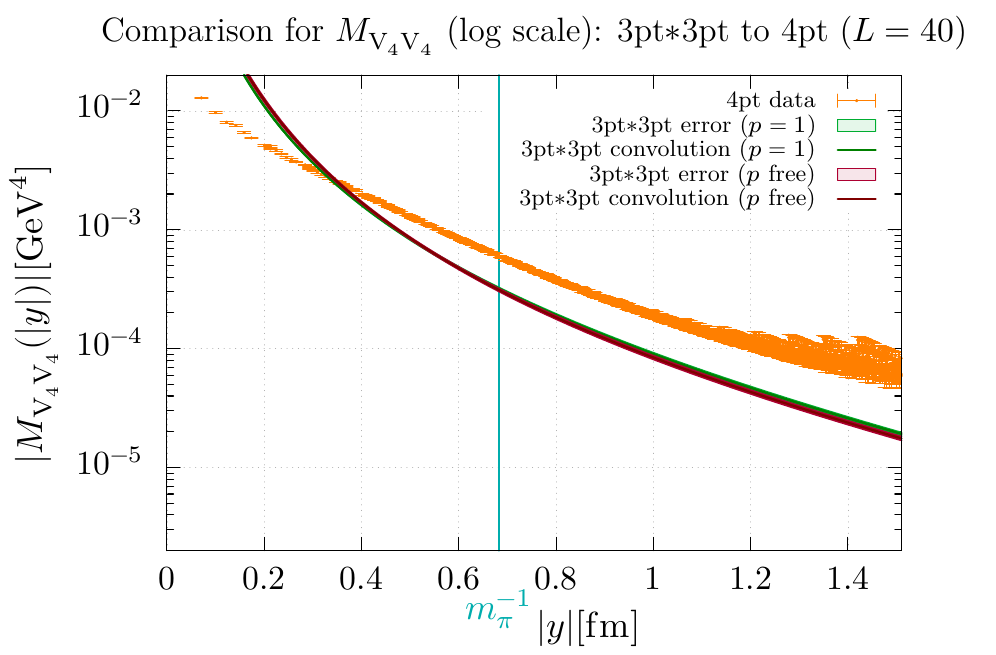}
\end{minipage}
\begin{minipage}[c][4cm]{7cm}
\includegraphics[width=7cm,trim={300 0 0 0},clip=true]{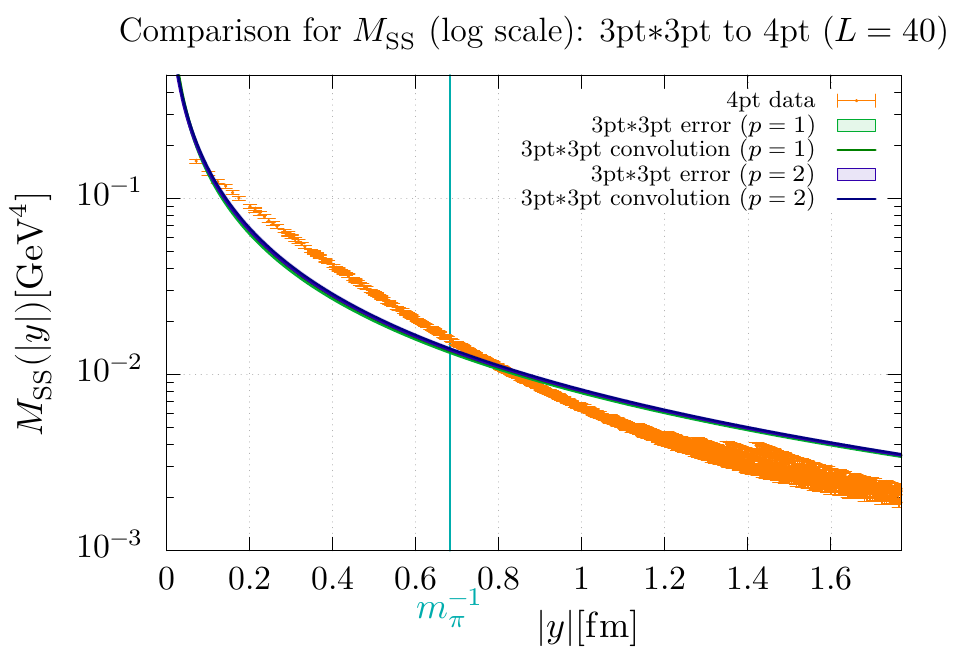}
\end{minipage}
\end{center}
\caption{Results for the convolution \protect\eqref{ftest2} compared with the 4pt data of the two-operator matrix elements for the SS and VV case (logarithmic scale).}
\label{fig:test2_res}
\end{figure}

The integrals \eqref{ftest1} and \eqref{ftest2} are evaluated numerically and compared to the corresponding 4pt data, where again only connected contributions are used. The results for test 1 are presented in \fref{fig:test1_res}. In the electromagnetic case one may observe a good agreement between the two curves for large distances, while for distances smaller than the pion wave length clear deviations are visible. For the scalar case a similar behavior can be found, although the agreement is slightly worse. This might be corrected by taking into account disconnected contributions, which are less suppressed in the scalar case.\\
The results of test 2 plotted in \fref{fig:test2_res} are very different. There is no region in $y$, where the 3pt data and the 4pt data show the same behavior. The naive factorization ansatz only predicts the correct order of magnitude for both, the electromagnetic and the scalar channel.

\section{Conclusion}
\label{sec:Conclusion}

We have investigated 4pt-functions, which are needed for the description of DPS processes, by performing lattice calculations. The calculations have been done for all 4pt-graphs, where we obtained good results for at least four graphs. For the remaining two graphs we still have to increase our statistics. We have checked the validity of the "naive factorization" assumption by performing two tests, where we have found that this kind of factorization fails for small parton distances, but also for large distances it might not be a valid assumption.\\
The efforts, which have been done so far, are planned to be extended to further aspects, like higher moments and non-zero momenta. Finally we want to investigate other particles, especially the nucleon, since this is the particle collided at LHC.

\providecommand{\href}[2]{#2}\begingroup\raggedright\endgroup

\end{document}